\documentstyle[preprint,aps]{revtex}
\begin{document}

\title{
\begin{flushright}
\vspace{-1cm}
{\normalsize MC/TH 95/22}
\vspace{1cm}
\end{flushright}
Determination of the pion-nucleon coupling constant\\
from QCD sum rules}
\author{Michael C. Birse and Boris Krippa\thanks{Permanent address: Institute 
for Nuclear Research of the Russian Academy of Science, Moscow Region 117312,
Russia.}}
\address{Theoretical Physics Group, Department of Physics and Astronomy\\
University of Manchester, M13 9PL, UK\\}
\maketitle
\begin{abstract}
We evaluate the $\pi N$ coupling constant using a QCD sum rule based on the
pion-to-vacuum matrix element of the correlator of two interpolating nucleon
fields. The part of the correlator with Dirac structure $k\llap/\gamma_5$ is
used, keeping all terms up to dimension 5 in the OPE and including continuum
contributions on the phenomenological side. The ratio of this sum rule to the
nucleon sum rule involving condensates of odd dimension yields stable results
with values of $g_{\pi N}$ in the range $12\pm 5$. The sources of uncertainty
are discussed.
\end{abstract}
\bigskip

Understanding hadron interactions from first principles is one of the main
goals of Quantum Chromodynamics (QCD). Since the solution of QCD for hadron
interactions at low energies is still far off it is useful to consider tackle
problems of hadron dynamics with approaches that lie as close as possible to
QCD. One of them, the method of QCD sum rules\cite{svz79}, has proved
to be a very powerful tool to extract information about hadron properties.

We present here a sum rule analysis of the $\pi N$ coupling constant $g_{\pi
N}$, which is one of the most important parameters in hadron physics. This
quantity has previously been examined within the framework of QCD sum rules in
Refs.~\cite{rry85,sh95}. Reinders, Rubinstein and Yazaki \cite{rry85} worked
with two different sum rules for $g_{\pi N}$, one obtained from a correlator
of three interpolating fields, and one based on the pion-to-vacuum matrix
element of the correlator of two interpolating nucleon fields, $\eta$:
\begin{equation}
\langle0|T\{\eta(x)\overline{\eta}(0)\}|\pi^a(k)\rangle,
\label{twopt}
\end{equation}
However those authors included only the leading term of operator product
expansion (OPE) and neglected continuum contributions. Shiomi and
Hatsuda\cite{sh95} developed the sum rule based on the part of this two-point 
correlation function with Dirac structure $\gamma_5$. They calculated $g_{\pi
N}$ in the soft-pion limit, including condensates up to dimension 7 in the
OPE, radiative corrections and a perturbative estimate of continuum
contributions.

The method based on the two-point correlator (\ref{twopt}) has a significant
advantage in that it can be used for low values of the momentum transfer to
the nucleon. In contrast the OPE of the three-point correlator is valid only
for large spacelike meson momenta and so a determination of the coupling
constant requires an extrapolation to zero momentum (see, for
example\cite{mei95}). This procedure is dangerous because of the higher-order
terms that have been omitted from the OPE. The contributions of these 
terms give rise to corrections that are proportional to large inverse powers 
of the meson momentum $k$, making the extrapolation of a truncated OPE 
unreliable. Estimates of the coupling constant from the coefficient of $1/k^2$
determined at large $k^2$, as in Refs.\cite{rry83,rry85}, cannot distinguish
the meson pole term from the contributions of higher-mass states in the same 
meson channel.

In the soft-pion limit the OPE for the $\gamma_5$ part of the two-point 
correlator for $g_{\pi N}$ has exactly the same form as that for the nucleon
sum rule\cite{iof81,is84} involving condensates of odd dimension, up a factor
of $1/f_\pi$ \cite{rry85,sh95}. If continuum corrections are neglected, the
ratio of these two sum rules has the form of the Goldberger-Treiman relation
with $g_A=1$\cite{rry85}. Shiomi and Hatsuda\cite{sh95} showed that this
continues to hold for the higher-dimension terms in the OPE, provided that the
continuum thresholds are taken to be the same for both sum rules. Using
different thresholds in the two sum rules they were able to get around this
problem with the implied value of $g_A$.

However, using the usual soft-pion theorem\cite{dgh92}, the correlator (1)
can be expressed as
\begin{equation}
-{i\over f_\pi}\langle 0|[Q_5^a, T(\eta(x),\eta^\dagger(0))]|0\rangle
={i\over 2f_\pi}\{\gamma_5\tau^a,\langle 0|T(\eta(x),
\eta^\dagger(0))|0\rangle\}
\end{equation}
where $Q_5^a$ is the axial charge and we have made use of the transformation
properties of the interpolating field under axial rotations\cite{lccg95},
$[Q_5^a,\eta]=-{1\over 2}\gamma_5\tau^a\eta$. The anticommutator with
$\gamma_5$ picks out the part of the two-point correlator proportional to the
unit Dirac matrix. The phenomenological side of the resulting sum rule is thus
$i\gamma_5/f_\pi$ times the corresponding expression for the odd-condensate
nucleon sum rule. This matches exactly with the structure found for the OPE
side in Refs.\cite{rry85,sh95}.

The soft-pion limit for the $\gamma_5$ piece of the correlator (1) thus yields
a sum rule for $M_N/f_\pi=g_{\pi N}/g_A$. The value for $g_{\pi N}$ determined
from such a rum rule follows from the odd-condensate sum rule for the nucleon
mass and the Goldberger-Treiman relation (or an approximation to it taking
$g_A=1$). The sum rule can be thought of as just a chiral rotation of the
odd-condensate nucleon sum rule and {\it not} an independent determination of
$g_{\pi N}$. By considering terms beyond the soft-pion limit, we obtain here a
value for $g_{\pi N}$ that is not simply a consequence of chiral symmetry. The
sum rule we use is thus analogous to that for the $\pi N\Delta$
coupling\cite{rry85} and the approach can be applied to calculations of other
meson-baryon couplings.

Moreover, a potentially important piece of the phenomenological side of
the sum rule for $g_{\pi N}$ has been omitted in both calculations. This term
corresponds to transitions of where a ground-state nucleon absorbs the pion
and is excited into the continuum. Since it is not suppressed by the Borel
transformation such a term should be included in a consistent sum-rule
analysis, as pointed out long ago\cite{is84,bk83} and stressed recently by
Ioffe\cite{iof95a,iof95b}. In the soft-pion limit, such terms generate contact
interactions where the pion couples directly to the nucleon field, $\langle
N(p)|\overline{\eta}_{n}(0)|\pi(k)\rangle$, and which are essential if the
correct soft-pion limit is to be obtained. Although the need for these terms
is particularly clear if pseudovector $\pi N$ coupling is used (see for
example\cite{koi93}), they should also be included for pseudoscalar coupling.
Their omission in Refs.\cite{rry85,sh95} can explain why the correct
Goldberger-Treiman relation was not found there. Indeed, as the authors
of\cite{sh95} point out, a quick estimate of these unsuppressed $N^*$
contributions suggests that they could be as large as 25\%: enough to remove
the discrepancy with the Goldberger-Treiman relation.

Here we start from the two-point correlator (\ref{twopt}) just discussed, but
instead of the piece with with Dirac structure $\gamma_5$ considered in
Refs.\cite{rry85,sh95} we work with the structure $k\llap/\gamma_5$, where $k$
is the pion momentum. We work here to leading order in a chiral expansion,
neglecting higher-order terms in the pion momentum or current quark mass.

We consider the two-point correlation function
\begin{equation}
\Pi(p)=i\int d^4x\exp(ip\cdot x)\langle 0|T\{\eta_{p}(x)
\overline{\eta}_{n}(0)\}|\pi^+(k)\rangle,    
\label{correl}
\end{equation}
where we use the Ioffe interpolating field\cite{iof81} for the proton,
\begin{equation}
\eta_{p}(x)=\epsilon_{abc}[u^{a}(x)^TC\gamma_{\mu}u^{b}(x)]\gamma_{5}
\gamma^{\mu}d^{c}(x),
\end{equation}
where $a,b,c$ are the colour indices and $C$ is the charge conjugation matrix.
(The corresponding neutron field $\eta_n$ is obtained by interchanging $u$ and
$d$ quark fields.) More general choices of interpolating field are
possible, as discussed in detail by Leinweber\cite{lei95}. For the
odd-condensate nucleon sum rule, which we make use of in our determination of
$g_{\pi N}$, it turns out that the Ioffe field is close to the optimal one as
determined in Ref.\cite{lei95} and so we do not consider more general fields 
here.

In the deeply Euclidean region ($p^2$ large and negative) the OPE of the
product of two interpolating fields takes the following general form
\begin{equation}
i\int d^4x\exp(ip\cdot x)T\{\eta_{p}(x)\overline{\eta}_{n}(0)\}
=\sum_n C_n(p)O_{n},
\end{equation}
where $C_{n}(p)$ are the Wilson coefficients and $O_{n}$ are local operators
constructed out of quark and gluon fields (all renormalised at some scale
$\mu$). Using this OPE in the correlator (\ref{correl}), we find that only
operators of odd dimension contribute. The leading term in this expansion
involves operators with dimension 3 and has the form
\begin{equation}
\Pi_{3}(p,k)=-\frac{1}{2\pi^{2}}p^{2}\ln(-p^{2})\langle 0|\overline{d}
\gamma^{\alpha}\gamma_{5}u|\pi^+(k)\rangle\gamma_{\alpha}\gamma_{5}+\cdots,
\end{equation}
where terms that do not contribute to the Dirac structure of interest, 
$k\llap/\gamma_5$, have been suppressed. The matrix element here is just 
the usual one for pion decay:
\begin{equation}
\langle 0|\overline{d}\gamma^{\alpha}\gamma_{5}u|\pi^+(k)\rangle=i\sqrt{2}
f_\pi k^\alpha,
\end{equation}
where $f_\pi=93$ MeV is the pion decay constant.

At dimension 5 the only relevant contribution arises from the second-order term
in the covariant expansion of the nonlocal operator $\overline{d}(0)
\gamma^{\alpha}\gamma_{5}u(x)$. It has the form
\begin{equation}
\Pi_{5}(p)=\frac{5}{9\pi^{2}}\ln(-p^{2})\langle 0|\overline{d}\gamma^{\alpha}
\gamma_{5}D^{2}u|\pi^+(k)\rangle\gamma_{\alpha}\gamma_{5}+\cdots.
\label{dim5}
\end{equation}
Up to corrections of higher order in the current mass, this can easily be
re-expressed in terms of a mixed quark-gluon condensate
\begin{equation}
\langle 0|\overline{d}\gamma^{\alpha}\gamma_{5}D^{2}u|\pi^+(k)\rangle=
\frac{g_{s}}{2}\langle 0|\overline{d}\gamma^{\alpha}\gamma_{5}\sigma_{\mu\nu}
G^{\mu\nu}u|\pi^+(k)\rangle+{\cal O}(m_c^2).
\end{equation}
Some further manipulation allows one to rewrite this in the form
\begin{equation}
\langle 0|\overline{d}\gamma^{\alpha}\gamma_{5}D^{2}u|\pi^+(k)\rangle=
-g_{s}\big(\langle 0|\overline{d}\widetilde{G}^{\alpha\mu}\gamma_{\mu}u
|\pi^+(k)\rangle-ig_s\langle 0|\overline{d}G^{\mu\alpha}\gamma_{\mu}\gamma_{5}u
|\pi^+(k)\rangle\bigr),
\label{cond5}
\end{equation}
where $\widetilde{G}_{\mu\nu}=\frac{1}{2}\epsilon_{\mu\nu\rho\sigma}
G^{\rho\sigma}$. (We use the convention $\epsilon^{0123}=+1$.) The second term
in this expression is of higher order in the chiral expansion (see
Ref.\cite{nsvvz} for details) and so we neglect it here.

The first matrix element in (\ref{cond5}) was extracted by Novikov {\it et
al.}\cite{nsvvz} from two QCD sum rules for the pion. They expressed it in the
form
\begin{equation}
g_{s}\langle 0|\overline{d}\widetilde{G}^{\alpha\mu}\gamma_{\mu}u
|\pi^+(k)\rangle=\sqrt{2}i\delta^{2}f_{\pi}k^{\alpha},
\label{delta2}
\end{equation}
and obtained $\delta^2=(0.20\pm 0.02)$ GeV$^2$.\footnote{There is a potential
sign ambiguity in using the result of Ref.\cite{nsvvz} since they do not
specify their convention for $\epsilon^{0123}$. However we have checked that
our coefficient of $x^2$ in the expansion of $\langle 0|\overline{d}(0)
\gamma^{\alpha}\gamma_{5}u(x)|\pi^+\rangle$ (which should be independent of
convention) agrees in both sign and magnitude with that of the corresponding
term in the expansion of $\langle 0|u(x)\overline u(0)|0\rangle_A$ given in
Refs.\cite{bk83,iof95b}.} A crucial contribution in both of their sum rules is
the four-quark condensate, $\alpha_s\langle 0|(\overline qq)^2 |0 \rangle$.
Novikov {\it et al.}\cite{nsvvz} used the factorisation approximation for this
quantity but direct determinations of it from other sum rules suggest
significantly larger values\cite{kar92,bnp92,nar95}, at least 2--3 times
bigger than those obtained from factorisation. These give correspondingly
larger values for $\delta^2$, a point we shall come back to in the discussion
of our results below.

As an estimate of the importance of higher dimension condensates, we have also
calculated the contribution of what we hope is the most important dimension-7
operator in the OPE. This is a mixed quark-gluon condensate, which we evaluate
in the factorised approximation. Keeping only this contribution explicitly,
the dimension-7 piece of the correlator is
\begin{equation}
\Pi_{7}(p)=-\frac{1}{12p^{2}}\langle 0|\overline{d}\gamma^{\alpha}\gamma_{5}u
|\pi^+(k)\rangle\langle 0|\frac{\alpha_{s}}{\pi}G^{2}|0\rangle\gamma_{\alpha}
\gamma_{5}+\cdots,
\end{equation}
where $\langle 0|\frac{\alpha_{s}}{\pi}G^{2}|0\rangle$ is the gluon condensate
in vacuum.

On the phenomenological side, the term of interest in the correlator
(\ref{correl}) is the one with a double pole at the nucleon mass, since this 
contains the $\pi N$ coupling constant. However the nucleon interpolating
field does not just create ground-state nucleons; there are also continuum
contributions which cannot be ignored. The continuum-to-continuum pieces are
modelled in the usual manner, in terms of the spectral density associated with
the imaginary part of the OPE expression for the correlator. This continuum is
assumed to start at some threshold $S_{\pi N}$. After Borel transformation, it
can be taken over to the OPE side of the sum rule where it modifies the
coefficients of the terms involving $\ln(-p^2)$. In addition one must include
nucleon-to-continuum terms since Borel transformation does not suppress these
with respect to the double-pole term\cite{is84,bk83,iof95a,iof95b}. To first
order in $k$, the correlator has the form
\begin{equation}
\Pi(p)=i{\sqrt 2}k\llap/\gamma_5\left[{\lambda_N^2 M_Ng_{\pi N}\over 
(p^2-M_N^2)^2}+\int_{W^2}^\infty ds\,b(s){1\over s-M_N^2}
\left({1\over p^2-M_N^2}+{a(s)\over s-p^2}\right)\right]+\cdots,
\label{phenom}
\end{equation}
where the continuum-continuum terms (and terms with other Dirac structures)
have not been written out. Here $\lambda_N$ is the strength with which the
interpolating field couples to the nucleon:
\begin{equation}
\langle 0|\eta_N(0)|N(p)\rangle=\lambda_N u({\hbox{p}}).
\end{equation}
Note that the strength of the $k\llap/\gamma_5$ piece of the double-pole term
is the same for both pseudoscalar and pseudovector $\pi N$ coupling.

Equating the OPE and phenomenological expressions for the correlator 
(\ref{correl}) and Borel transforming\cite{svz79}, we get the sum rule
\begin{equation}
\frac {1}{2\pi^{2}}M^4E_{2}(x) + \frac{5}{9\pi^{2}}M^2E_{1}(x)\delta^{2}
+\frac{1}{12}E_{0}(x)\langle 0|\frac{\alpha_{s}}{\pi}G^{2}|0\rangle
=\left({\lambda_N^2 M_N g_{\pi N}\over f_{\pi}M^2}+A\right)
\exp(-M^{2}_{N}/M^{2}),
\label{gpinsr}
\end{equation}
where $M$ is the Borel mass and $E_{n}(x) = 1 - (1 + x +...+\frac{x^{n}}{n!})
e^{-x}$ with $x = \frac{S_{\pi N}}{M^{2}}$. The second term on the r.h.s.~of
this sum rule, involving the undetermined constant $A$, is the Borel transform
of the nucleon pole term of the nucleon-to-continuum piece in (\ref{phenom}).
It contains the same exponential as the nucleon double-pole term and so cannot
be ignored. The second nucleon-to-continuum term in (\ref{phenom}) leads to a
term that is suppressed by an exponential involving the masses of states in
the continuum. It is thus typically a factor of 3--4 smaller than the term
included in (\ref{gpinsr}). Provided that the first of these mixed terms is a
reasonably small correction to the sum rule, it should be safe to neglect the
second, as discussed by Ioffe\cite{iof95a,iof95b}.

We now turn to the numerical analysis of this sum rule. First, one should get
rid of the unknown constant $A$. Multiplying the sum rule by
$M^2\exp{M_N^2/M^2}$, we see that the r.h.s.~becomes a linear function of
$M^2$. By acting on this form of the sum rule with $(1-M^2\partial/\partial
M^2)$\cite{is84} (or equivalently by fitting a straight line to the l.h.s.~and
extrapolating to $M^2=0$\cite{bk83}) we in principle can determine the value
of $g_{\pi N}$. However we are unable to find a region of Borel mass in which
the l.h.s.~is approximately a linear function of $M^2$, and hence there is no
region of stability for the extracted $g_{\pi N}$.

This lack of stability is similar to the situation for the nucleon sum rules,
where two sum rules can be derived\cite{iof81} (involving either odd or even
dimension operators) but neither shows good stability. Nonetheless the ratio of
these leads to a more stable expression for the nucleon mass. We have therefore
taken the ratio of our sum rule (\ref{gpinsr}) to the nucleon sum rules. We
obtain the most stable results from the ratio to the odd-dimension sum rule,
\begin{equation}
-{1\over 4\pi^2}M^4E_1(x_N)\langle 0|\overline q q|0\rangle+{1\over 24}
\langle 0|\overline q q|0\rangle\langle 0|\frac{\alpha_{s}}{\pi}G^{2}|0\rangle
=\lambda_N^2 M_N \exp(-M^{2}_{N}/M^{2}),
\label{noddsr}
\end{equation}
and so we present here only the results for that case. Taking such a ratio also
has the advantage of eliminating the experimentally undetermined strength
$\lambda_N$ from the sum rules. Note that we have allowed for a different
continuum threshold $S_N$ in the nucleon sum rule and have defined 
$x_N=S_N/M^2$.

The ratio of the sum rules (\ref{gpinsr}) and (\ref{noddsr}) can be written in
the form
\begin{equation}
f_\pi{\frac {1}{2\pi^{2}}M^6E_{2}(x) + \frac{5}{9\pi^{2}}M^4E_{1}(x)\delta^{2}
+\frac{1}{12}M^2E_{0}(x)\langle 0|\frac{\alpha_{s}}{\pi}G^{2}|0\rangle\over
-{1\over 4\pi^2}M^4E_1(x_N)\langle 0|\overline q q|0\rangle+{1\over 24}
\langle 0|\overline q q|0\rangle\langle 0|\frac{\alpha_{s}}{\pi}G^{2}|0\rangle}
=g_{\pi N}+A'M^2,
\label{ratio}
\end{equation}
and the method discussed above used to eliminate the unknown mixed
nucleon-to-continuum term, $A'M^2$ ($A'=Af_\pi/\lambda_N^2M_N$). In Fig.~1 we
show results for $g_{\pi N}$ as a function of $M^2$, for typical values of the
condensates and thresholds:
$\langle 0|\overline q q|0\rangle=-(0.245\ {\hbox{GeV}})^3$, 
$\langle 0|\frac{\alpha_{s}}{\pi}G^{2}|0\rangle\simeq 0.012$ GeV$^{4}$,
$\delta^2=0.35$ GeV$^2$, $S_N=2.5$ GeV and $S_{\pi N}=2.15$ GeV. Stable 
values of $g_{\pi N}\simeq 11.7$ are found over a region $M^2\simeq 0.8-1.8$ 
GeV$^2$. Corrections due to the $A'M^2$ term are small, at most 5\%. The
second such term in (\ref{phenom}) is expected to be smaller by a factor of
3--4, and so we are justified in neglecting it.

The threshold $S_{\pi N}$ has been adjusted to give stability for $M^2$ around
1 GeV$^2$, since one may hope that in this region the Borel transformed sum
rule is not too sensitive to the approximations that have been made on both
the OPE and phenomenological sides of the sum rule. The existence of a window
of stability provides a check on the consistency of this assumption. We also
demand that the thresholds $S_N$ and $S_{\pi N}$ should lie significantly
above this window so that the continuum is not too heavily weighted in the
Borel transform. We find that the window of stability moves rapidly upwards as
$S_{\pi N}$ is increased for fixed $S_N$. For the typical parameter values
above, only the region 2.05 GeV$^2\leq S_{\pi N}\leq 2.22$ GeV$^2$ satisfies
these requirements. The value of $g_{\pi N}$ varies by at most $\pm 0.2$ over
this region.

As a further check on our results, we have examined whether the individual sum
rules (\ref{gpinsr}) and (\ref{noddsr}) satisfy the criteria suggested by
Leinweber\cite{lei95}. We find that the highest dimension condensates
contribute less that 10\% of the OPE to both sum rules for $M^2>0.8$ GeV$^2$.
The continuum forms about 40\% of the phenomenological side of the
differentiated version of the $g_{\pi N}$ sum rule (\ref{gpinsr}) for $M^2$ up
to 1.4 GeV$^2$, the point at which the continuum reaches 50\% of the
odd-condensate sum rule (\ref{noddsr}). The region $M^2\simeq 0.8-1.4$ GeV$^2$
thus provides a window within which our results are both stable with respect
to the Borel mass and not too sensitive to our approximations.

We have examined the dependence of our results on the other input parameters.
Variation of the threshold in the nucleon sum rule $S_N$ from 2.2 to 2.8
GeV$^2$, readjusting $S_{\pi N}$ to maintain stability, changes $g_{\pi N}$ by
$\pm 0.2$. To estimate the sensitivity of our sum rule to the contributions of
dimension-7 condensates and to uncertainties in the gluon condensate, we have
varied the dimension-7 term in (\ref{gpinsr}) between zero and twice its
standard value. Our results for $g_{\pi N}$ change by $\pm 0.5$ over this
range.

One of the most important input parameters in our sum rule is the matrix
element $\delta^2$, defined by (\ref{delta2}). As already mentioned, this
parameter was extracted by Novikov {\it et al.}\cite{nsvvz} from an analysis
of two sum rules for the pion. Their results depend crucially on the
four-quark condensate, $\alpha_s\langle 0|(\overline qq)^2 |0 \rangle$, for
which they made the factorisation approximation and took a value of about
$2\times 10^{-4}$ GeV$^6$. With this input, both of their sum rules yield
consistent results for $\delta^2$ in the region $0.20 \pm 0.02$ GeV$^2$.
However, sum rules analyses of $\tau$ decay and $e^+e^-$ annihilation into
hadrons lead to significantly larger values of the four-quark condensate
(see\cite{kar92,bnp92,nar95} and references therein), in the range
$(4-6)\times 10^{-4}$ GeV$^6$. Using these in the sum rules of
Ref.\cite{nsvvz} leads to values for $\delta^2$ ranging from 0.28 to 0.45,
although the two sum rules do not then give consistent results. As a
conservative estimate of the uncertainty in $\delta^2$ we have considered the
range 0.20 to 0.45 GeV$^2$. The corresponding variation in $g_{\pi N}$ is $\pm
2$, when the other parameters are held at their values above and $S_{\pi N}$
is changed to keep the window of stability around 1 GeV$^2$.

A second significant source of uncertainty is the quark condensate 
$\langle 0|\overline q q|0\rangle$ which appears in the odd-dimension sum
rule for the nucleon. ``Standard" values for this lie in the range $-(0.21\
{\hbox{GeV}})^3$ and $-(0.26\ {\hbox{GeV}})^3$. The values of the nucleon mass
determined from sum rules\cite{iof81} are strongly correlated with this
condensate. There is also a weaker correlation with the chosen value of the
threshold $S_N$. Since we are dividing by $M_N$ in the ratio (\ref{ratio}),
our results are rather sensitive to the value of this condensate. One would
like to use values of $\langle 0|\overline q q|0\rangle$ and $S_N$ that give
the nucleon mass correctly, but the ratio of the odd and even dimension nucleon
sum rules does not yield completely stable results for $M_N$. The best we can
do is to rule out values of $-\langle 0|\overline q q|0\rangle$ below $(0.23\
{\hbox{GeV}})^3$ since they cannot reproduce the nucleon mass within the
region of Borel mass and threshold that we consider. Varying the quark
condensate between $-(0.23\ {\hbox{GeV}})^3$ and $-(0.26\ {\hbox{GeV}})^3$, we
find that $g_{\pi N}$ changes by $\pm 2$.

Our final result for the $\pi N$ coupling constant is thus $g_{\pi N}=12\pm 5$,
where the uncertainty is dominated by $\delta^2$ and $\langle 0|\overline q
q|0\rangle$. This is to be compared with values deduced from $NN$ and $\pi N$
scattering. For many years the accepted value was $g_{\pi N}=13.4$\cite{bcc73}
but this coupling has been the subject of some debate in recent years. More
recent analyses lead to values in the range 12.7--13.6\cite{newgpi}. Our result
is obviously consistent with any of these. The rather large uncertainty in it
could be reduced if the quark condensate could be determined more precisely.
In addition, the sum rules of Novikov {\it et al.}\cite{nsvvz} should be
re-examined using larger values of the four-quark condensate to try to pin
down $\delta^2$ more exactly.

In summary, we have calculated the $\pi N$ coupling constant using a QCD sum
rule based on the pion-to-vacuum matrix element of a two-point correlator
of interpolating nucleon fields. This approach avoids the need for
extrapolation from large spacelike meson momenta. We have included
nucleon-to-continuum terms omitted from previous analyses. Our sum rule is
based on the part of the correlator with Dirac structure $k\llap/\gamma_5$ and
includes all terms up to dimension 5 in the OPE. Stable results are obtained
from the ratio of this sum rule to one for the nucleon mass and the
unsuppressed nucleon-to-continuum contributions are found to be small.
Contributions from higher-dimension operators and omitted continuum
contributions are estimated to be small. This demonstrates the practicability
of this type of sum rule for calculation of other meson-baryon couplings,
whose values are at present not well determined.

\section*{Acknowledgements}
We are grateful to V. Kartvelishvili for useful discussions. This work was 
supported by the EPSRC and PPARC.

\newpage
\begin{center}
{\large FIGURE CAPTION}
\end{center}
\bigskip
Fig.~1. Dependence on the square of the Borel mass of the $\pi N$ coupling
constant determined from the ratio of sum rules (\ref{ratio}). The values of
the parameters used are given in the text. The solid line shows the value of
$g_{\pi N}$ corrected for the mixed continuum term $A'M^2$, the dashed line the
uncorrected l.h.s.~of (\ref{ratio}).


\begin{thebibliography}{999}
\bibitem{svz79}M. A. Shifman, A. I. Vainshtein and V. I. Zakharov, 
Nucl.\ Phys.\ {\bf B147} (1979) 385, 448.
\bibitem{rry85}L. J. Reinders, H. Rubinstein and S. Yazaki, Phys.\ Reports 
{\bf 127} (1985) 1.
\bibitem{sh95}H. Shiomi and T. Hatsuda, Nucl.\ Phys.\ {\bf A594} (1995) 294.
\bibitem{mei95}T. Meissner, Phys.\ Rev.\ {\bf C52} (1995) 3386.
\bibitem{rry83}L. J. Reinders, H. Rubinstein and S. Yazaki, Nucl.\ Phys.\ {\bf
 B213} (1983) 109.
\bibitem{iof81}B. L. Ioffe, Nucl.\ Phys.\ {\bf B188} (1981) 317; {\bf B191} 
591(E).
\bibitem{is84}B. L. Ioffe and A. V. Smilga, Nucl.\ Phys.\ {\bf B232} (1984) 
109.
\bibitem{dgh92}J. F. Donoghue, E. Golowich and B. R. Holstein, {\it Dynamics
of the standard model} (Cambridge University Press, Cambridge, 1992).
\bibitem{lccg95}S. H. Lee, S. Cho, T. D. Cohen and D. K. Griegel,
Phys.\ Lett.\ {\bf B348} (1995) 263.
\bibitem{bk83}V. M. Belaev and Ya. I. Kogan, Phys.\ Lett.\ {\bf 136B} (1983) 
273. 
\bibitem{iof95a}B. L. Ioffe, University of Bern preprint BUTP-94/25, 
hep-ph/9501319 (1995).
\bibitem{iof95b}B. L. Ioffe, ITEP preprint 62-95, hep-ph/9511401 (1995).
\bibitem{koi93}Y. Koike, Phys.\ Rev.\ {\bf D48} (1993) 2313.
\bibitem{lei95}D. B. Leinweber, University of Washington preprint 
DOE/ER/40427-17-N95, nucl-th/9510051.
\bibitem{nsvvz}V. A. Novikov, M. A. Shifman, A. I. Vainshtein, M. B. Voloshin 
and V. I. Zakharov, Nucl.\ Phys.\ {\bf B237} (1984) 525. 
\bibitem{kar92}V. G. Kartvelishvili and M. V. Margvelashvili, Z. Phys.\ {\bf
C55} (1992) 83;
V. Kartvelishvili, Phys.\ Lett.\ {\bf B287} (1992) 159.
\bibitem{bnp92}E. Braaten, S. Narison and A.Pich, Nucl.\ Phys.\ {\bf B373} 
(1992) 581.
\bibitem{nar95}S. Narison, Phys.\ Lett.\ {\bf B361} (1995) 121.
\bibitem{bcc73}D. V. Bugg, A. A. Carter and J. R. Carter, Phys.\ Lett.\ {\bf 
44B} (1973) 278.
\bibitem{newgpi}V. Stoks, R. Timmermans and J. J. de Swart, Phys.\ Rev.\ {\bf
C47} (1993) 512;\\
R. A. Arndt, R. L. Workman and M. Pavan, Phys.\ Rev.\ {\bf C49} (1994) 2729;\\
F. Bradamante, A. Bressan, M. Lamanna and A. Martin, Phys.\ Lett.\ {\bf B343}
(1995) 431;\\
T. E. O. Ericson {\it et al.}, Phys.\ Rev.\ Lett.\ {\bf 75} (1995) 1046;\\
D. V. Bugg and R. Machleidt, Phys.\ Rev.\ {\bf C52} (1995) 1203.

\end{thebibliography}
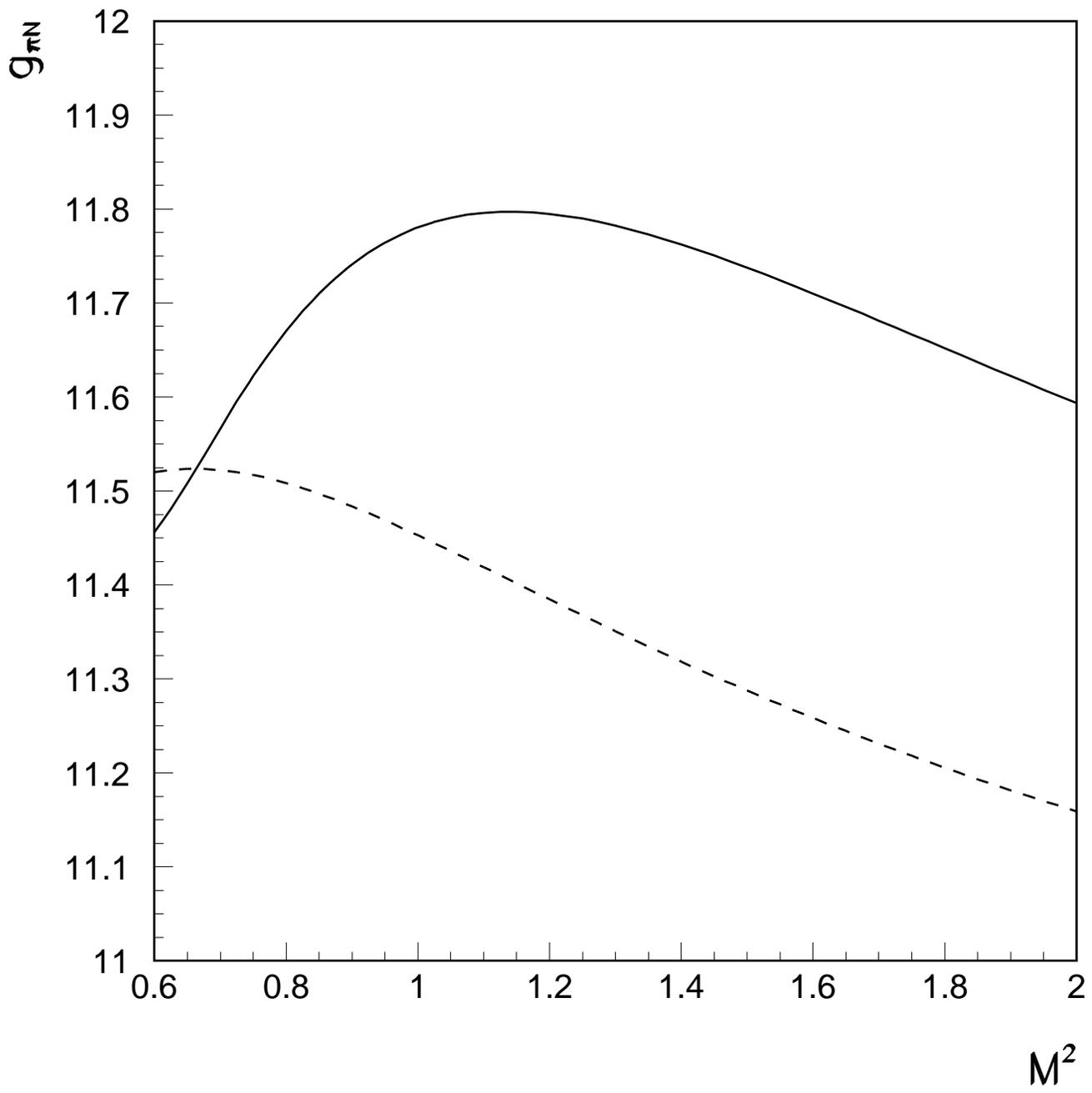
\end{document}